\documentclass[prc,showpacs,letterpaper,aps]{revtex4}

\usepackage{graphicx}

\begin{document}

\title{Extension of the Random Phase Approximation including the 
self-consistent coupling to two-phonon contributions}
\author{C.~Barbieri}
  \email{barbieri@triumf.ca}
  \homepage{http://www.triumf.ca/people/barbieri}
  \affiliation
     {TRIUMF, 4004 Wesbrook Mall, Vancouver, 
          British Columbia, Canada V6T 2A3 \\  }

\author{W.~H.~Dickhoff}
  \email{wimd@wuphys.wustl.edu}
  \homepage{http://www.physics.wustl.edu/~wimd}
  \affiliation
     {Department of Physics, Washington University,
	 St.Louis, Missouri 63130, USA \\  }

\date{\today}

\begin{abstract}
A microscopic formalism is developed that includes the coupling
to two particle-hole phonons in the particle-hole propagator
by extending the dressed random phase approximation
(DRPA) equation for a finite system.
 The resulting formalism is applied to study the low-lying
excitation spectrum of ${}^{16}\textrm{O}$.
 It is observed that the coupling to two-phonon states at low energy
generates excited states with
quantum numbers that cannot be obtained in the DRPA approach.
 Nevertheless, the two-phonon states mix weakly with particle-hole
configurations and participate only partially in the formation of 
the lowest-lying positive-parity excited states.
 The stability of the present calculation is tested vs. the truncation of
model space. 
It is demonstrated that when single-particle strength
fragmentation is properly considered, the present formalism 
exhibits convergence with respect to the chosen model space
within the confines of the chosen approximation scheme.
\end{abstract}

\maketitle

\section{Introduction}
\label{sec:1}

Recent interest in the spectroscopic factors for the removal of low-lying
$p_{1/2}$ and $p_{3/2}$ strength from ${}^{16}\textrm{O}$ points to
a substantial discrepancy between experiment~\cite{leus}
and theory~\cite{md94,rad94,pmd95,muetherO16,GeurtsO16,O16jast,Fabr01,
FaddRPAO16,muether02,udias,GiustiG,WimMarco}.
The contribution to the reduction of these spectroscopic factors due
to short-range correlations is important and is uniformly calculated
to be about 10\% in ${}^{16}\textrm{O}$ using different realistic 
interactions and theoretical approaches.
The best agreement with the data is obtained when also long-range correlations
are included in the
calculations~\cite{muetherO16,GeurtsO16,FaddRPAO16,muether02}.
The results of Ref.~\cite{FaddRPAO16} demonstrate a particularly
strong correlation between the theoretically calculated
excitation spectrum of ${}^{16}\textrm{O}$ and the resulting
one-hole spectral function,
which describes the excitations of the residual $A-1$ nucleus.
The appearance of an additional $p_{3/2}$ fragment at small missing
energy when a low-lying excited $0^+$ state is obtained is one such
correlation~\cite{FaddRPAO16}.
Another one is the appearance of low-lying $s_{1/2}$ and $d_{5/2}$
strength when the first $3^-$ state in ${}^{16}\textrm{O}$ is correctly
described~\cite{GeurtsO16,FaddRPAO16}.
Both features are in accordance with the experimental data~\cite{leus}.

The calculation of Ref.~\cite{FaddRPAO16} employed a Random Phase
Approximation (RPA) approach
to describe the excitation spectrum of ${}^{16}\textrm{O}$.
The quality of the spectrum in this approach
is, however, quite inadequate.
Even with realistic $G$-matrices as residual particle-hole (ph) interaction
one generates at most a few low-lying collective isoscalar states of negative
parity ($3^-$ and $1^-$) and no low-lying isoscalar positive parity
states are obtained~\cite{czer}.
 These considerations suggest that an adequate description of the
sp fragmentation, at low energy, requires a better description of the
excited states within the same framework of self-consistent Green's function
(SCGF).

 This improvement of the SCGF method is also important because it can be 
applied to heavier nuclei.
 This is possible since the method describes the features of the low-energy
spectrum in terms of interactions between a relatively small number of
quasiparticles and collective excitations.  These modes represent the
experimentally observed excitations of the system and therefore they include
both the effects of short- and long-range correlations.
 The fragmentation of the single-particle (sp) strength is thus
self-consistently included.
Early applications to  ${}^{48}\textrm{Ca}$ and ${}^{90}\textrm{Zr}$
have been reported in Refs.~\cite{brand88,brand90}. These calculations 
did not include self-consistency but nevertheless were able to yield 
a reasonable description of giant resonances
and Gamow-Teller states as well as the low-lying collective states.
 Moreover a formalism based on the Bethe-Salpeter equation for
ph excitations can be extended to include
the effects of the continuum~\cite{continuum}.
 Although the SCGF method has only been applied in doubly closed-shell
nuclei, it is feasible that increased computational power will allow
applications to open-shell systems.
 An exploratory calculation for semi-magic isotopes of $\textrm{Ni}$
and $\textrm{Sn}$ was reported in~\cite{JieTh}.
In the near future a huge amount of data will become available for 
unstable nuclei from radioactive beam facilities.
For this reason it is important to develop techniques that are flexible
enough to describe all the above effects.

In this paper, we report a first step toward an extension of the SCGF
formalism in this direction. We chose to use ${}^{16}\textrm{O}$
as a test system for the following reasons. First, a successful calculation
for this nucleus will help resolving the aforementioned discrepancies for
the spectroscopic factors. Second, this system has been studied  by various
approaches in the past. From the results of these calculations one can infer
the relevant physical ingredients to be included in the SCGF method.
 Third, since the spectrum of ${}^{16}\textrm{O}$ is rather complicated,
meeting the challenge of describing this system will virtually guarantee
good results when the same method is applied to heavier nuclei.

Shell-model calculations for this nucleus have been reported in
Refs.~\cite{zbg68,rw73,Haxton0+,wbm92}
which indicate the particular importance of four-particle four-hole (4p4h)
admixtures to the 0p0h ground state in generating
the first excited $0^+$ state. Other positive parity states at low energy
are dominated by 2p2h components, as indirectly confirmed
by inelastic electron scattering~\cite{O16ph86}.
Some of this resulting physics was anticipated in terms of deformation
effects which simulate these types of many-particle many-hole 
admixtures~\cite{Brown0+}.
A complementary point of view is given by the Interacting Boson Model
of Refs.~\cite{iach1,iach2}. There, the low-lying positive parity states
are understood in terms of the coupling of $3^-$ and $1^-$ isoscalar states.
 One should note that, due to the predominant 1p1h nature of these
excitations, the coupling of different phonons generates the npnh
configurations relevant to this problem.
In Refs.~\cite{iach1,iach2} the $3^-$ and $1^-$ states were used as
phenomenological boson degrees of freedom to generate the
spectrum of the nucleus. 
 The connection with the underlying fermionic description was indicated
but not explored completely.
This relation was considered in Ref.~\cite{wim2ph}. There it was argued
that the microscopic
ph interaction contains two-phonon exchange contributions which
include the actually observed low-lying states of ${}^{16}\textrm{O}$
themselves, thereby generating the correct number of low-lying states
observed at low excitation energy.

In this work we begin to implement the physical ingredients
proposed in  Refs.~\cite{iach1,iach2} by extending the RPA to include
the coupling to two-phonon states.
 The basic idea of the present approach is illustrated in 
Fig.~\ref{fig:2PiERPA}.
This figure depicts the coupling of a ph state to two intermediate
phonons that are described by the ph propagator itself.
If the two intermediate phonons have been computed using the RPA
equations, they will already provide a reasonable description of the 
low-lying collective isoscalar $3^-$ and $1^-$ excited states.
 These phonons can be sufficient
to generate the quantum numbers of the most important positive parity states.
Particularly relevant is the coupling of two $3^-$ phonons,
$3^- \otimes 3^- = 0^+ \oplus 2^+ \oplus 4^+ \oplus 6^+$, that 
represent some  of the correlated 2p2h states of ${}^{16}\textrm{O}$.
 In the framework of SCGF one employs dressed sp propagator in the
construction of the microscopic phonons. This results in a Dressed RPA (DRPA)
approach for the ph calculation and in its extension to the coupling
to two-phonon states.
 In the present paper, we employ the self-consistent sp propagator
of ${}^{16}\textrm{O}$ computed in Refs.~\cite{FaddRPA1,FaddRPAO16}.
It should be noted that the incorporation of all the 4p4h effects in the
present formalism  requires a full four-phonon calculation.
 For this reason, a complete resolution in terms of a microscopic
description of the spectrum may only be partially successful.

 We note that there has been a tremendous progress in recent years in
the microscopic description of {\em p} shell nuclei using Green's Function
Monte Carlo and no-core shell model methods~\cite{GFMC,NoCshell,3Nforce}.
 A possible application of the no-core shell model to ${}^{16}\textrm{O}$
would properly include such 4p4h effects. However the description of
spectroscopic
factors would still require the construction of effective operators to 
include the effects of short-range correlations on these quantities
whereas these are automatically included in the SCGF method.

The paper is organized as follows: Sec.~\ref{sec:2PiERPA} introduces
the formalism to account for two-phonon coupling
in the calculation of the ph propagator.
 The approach employed here is based on a formalism first
introduced by Baym and Kadanoff for the description of response functions
in a many-body system at finite temperature~\cite{bk61,baym62,BKBook}.
 This framework provides a procedure to construct the effective interaction
of the ph Bethe-Salpeter equation that generalizes the (D)RPA approach.
 This method is described in Sec.~\ref{sec:BayKad} and the resulting equations
in Sec.~\ref{sec:ERPA_eq}.  More technical details are left
to the Appendix.
Sec.~\ref{sec:results} describes the results for the spectrum
within the current approximation scheme.
Sec.~\ref{sec:trunc} is devoted to a study of convergence properties
related to the number of two-phonon configurations included,
and the role of time-inversion diagrams. 
 In Sec.~\ref{sec:instab_0+} we discuss the possible appearance of
instabilities, in particular for the $0^+$ state, that was also observed
in Ref.~\cite{FaddRPAO16}.
Conclusions are drawn in Sec.~\ref{sec:concl}.

\section{Extension of \lowercase{ph}(D)RPA Formalism.}
\label{sec:2PiERPA}

In this work the central quantity of interest is the two-time polarization
propagator, whose Lehmann~\cite{lehmann} representation is given by
\begin{equation}
\Pi_{\alpha  \beta  , \gamma  \delta}(\omega) ~=~
 \sum_{n \neq 0} 
 \frac{\left( {\cal Z}^{n}_{\alpha \beta } \right)^* 
           \; {\cal Z}^{n}_{\gamma \delta}}
     {\omega - \varepsilon^{\pi}_n  + i \eta }
 ~-~ \sum_{n \neq 0} 
  \frac{  {\cal Z}^{n}_{\beta \alpha} 
           \;  \left( {\cal Z}^{n}_{\delta \gamma} \right)^* }
     {\omega + \varepsilon^{\pi}_n  - i \eta }  \; .
\label{eq:Pi}
\end{equation}
In Eq.~(\ref{eq:Pi}), the poles and residues contain the information on the
response and excitation energies of the system with $A$ particles in terms
of the quantities
\begin{eqnarray}
    {\cal Z}^{n}_{\alpha \beta} &=& 
      {\mbox{$\langle {\Psi^A_{n}} \vert $}}
        c^{\dag}_\alpha  c_\beta {\mbox{$\vert {\Psi^A_0} \rangle$}} \; ,       
\nonumber  \\
   \varepsilon^{\pi}_{n} &=& E^A_n - E^{A}_0  \; ,
\label{eq:PiAmpl_def}  
\end{eqnarray}
where $E^A_n$ and ${\mbox{$\vert {\Psi^A_n} \rangle$}}$  are the exact energies
and eigenstates of the $A$-particle system, the subscript $0$ refers to the 
ground state and $c^{\dag}_\alpha$~($c_\alpha$) is the creation~(annihilation)
operator of a particle in the state $\alpha$.
For a clearer discussion of the formalism employed in this work, it is useful
to first give a description of the relevant contributions in terms
of Feynman diagrams.

\subsection{Diagrammatic contributions}
\label{sec:BayKad}

\begin{figure}
 \begin{center}
\vspace{1.5in}
 \includegraphics[height=1.5in]{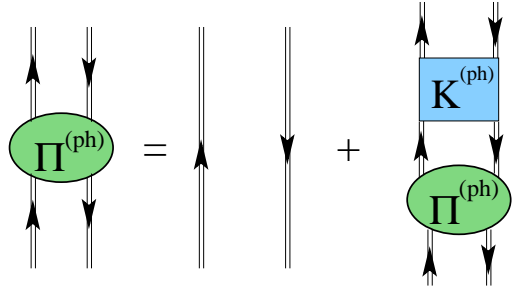}
 \end{center}
\caption[]{\small 
 Bethe-Salpeter equation for the ph polarization propagator. No 
specific time direction has to be assumed for these diagrams.
\label{fig:BS_eq} }
\end{figure}

\begin{figure}
 \begin{center}
\vspace{1.5in}
 \includegraphics[width=4in]{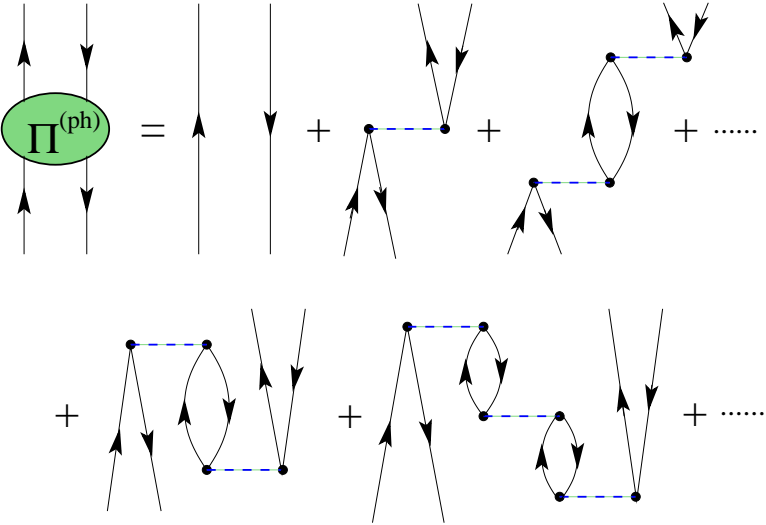}
 \end{center}
\caption[]{\small 
 Diagrammatic expansion of the standard RPA equation. An explicit time
 direction is assumed for the diagrams of this figure.
\label{fig:RPA_exp} }
\end{figure}

The exact ph propagator~(\ref{eq:Pi}) is a solution of the Bethe-Salpeter
equation, depicted in
Fig.~\ref{fig:BS_eq}~\cite{fetwa,AAA}. This equation
can be written schematically as
\begin{equation}
  \Pi ~ = ~ \Pi^{f} ~+~ \Pi^{f} ~ K^{(ph)} ~ \Pi
\label{eq:BS_eq}
\end{equation}
where $\Pi^{f}$ represents the free propagation of a quasiparticle
and a quasihole in the nuclear medium
and the ph kernel $K^{(ph)}$ is, in general, a four-time quantity.
According to the Baym-Kadanoff procedure, a solution for $\Pi$ is obtained
by first generating a self-consistent solution of the sp propagator
using a given choice of the self-energy. From the functional 
derivative of this (self-consistent) self-energy with respect to the
corresponding sp propagator, one then obtains the
irreducible ph interaction $K^{(ph)}$  that
generates  the corresponding conserving approximation for the ph propagator
(when used in the Bethe-Salpeter equation).
The standard RPA approach is derived by applying this procedure to the
the Hartree-Fock (HF) sp propagator and self-energy.
 This corresponds to  approximating $K^{(ph)}$ with the bare interaction
$V$ and employing bare (HF) sp propagators as external lines. 
Eq.~(\ref{eq:BS_eq}) then generates the RPA series of ring diagrams
shown in Fig.~\ref{fig:RPA_exp}.

\begin{figure}
 \begin{center}
\vspace{1.5in}
 \includegraphics[height=1.5in]{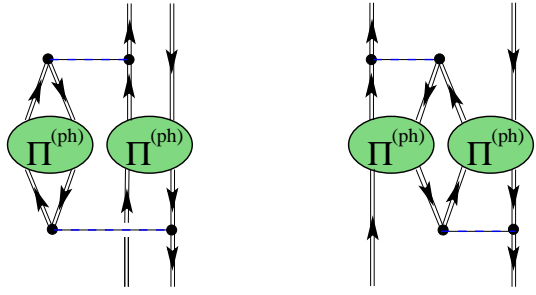}
 \end{center}
\caption[]{\small
 Examples of contributions involving the coupling of two independent
ph phonons. 
 In total, there are sixteen possible diagrams of this type, obtained
by considering all the possible couplings to a ph state.
The two-phonon ERPA equations~(\ref{eq:2PiERPA}) sum all of these
contributions in terms of dressed sp propagators.
\label{fig:2PiERPA} }
\end{figure}

\begin{figure}
 \begin{center}
\vspace{1.5in}
 \includegraphics[height=2in]{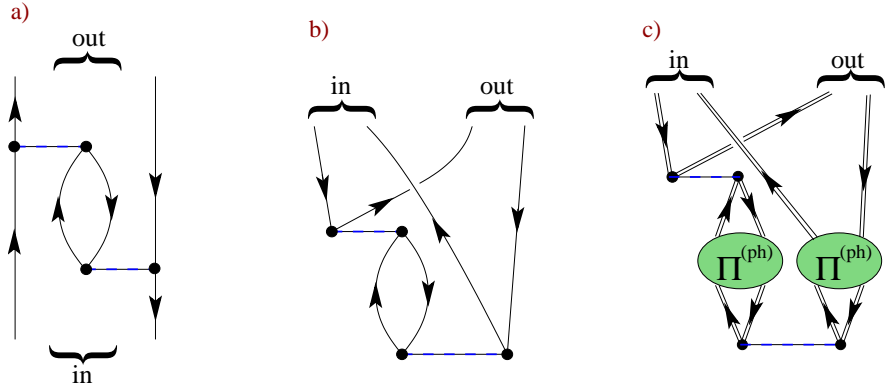}
 \end{center}
\caption[]{\small
 Example of direct, a), and time inversion, b), diagrams that appear in the
standard ERPA expansion. Both diagrams a) and b) come from the same four-time
screening diagram.
 The last picture in part c) shows the corresponding two-phonon extension
of the time-inversion contribution.
 Note that the diagram b) generates Pauli exchange corrections to the
 last diagram shown in Fig.~\ref{fig:RPA_exp}.
\label{fig:ERPAflip} }
\end{figure}

The extension of the RPA formalism, proposed and implemented 
by Brand \textit{et al.}~\cite{brand88,brand90},
was suggested by the observed fragmentation of the sp strength
and the necessity to go beyond a lowest-order self-energy for a commensurate
theoretical description.
There, the Baym-Kadanoff procedure is applied to a second-order approximation
for the self-energy. This yields contributions to the
kernel $K^{(ph)}$ that automatically include all the terms which
couple the ph states to the 2p2h ones, in accordance with the Pauli
principle for the latter states. Thus, this formalism
takes into account the mixing with 2p2h configurations
in the construction of the ph propagator
(the diagram of Fig.~\ref{fig:ERPAflip}a gives an example).
In the work of Brand \textit{et al.} only a single-pole approximation
to the self-consistent propagators was employed.
It is the aim of the present work to take the effects of the sp 
fragmentation more completely into account
and therefore a fully dressed sp propagator must be used.
The one employed in the present work
was obtained in Ref.~\cite{FaddRPAO16} by means 
of a Faddeev expansion for the nuclear self-energy. 
 When one applies the Baym-Kadanoff prescription to the latter self-energy,
a large set of contributions to $K^{(ph)}$ is generated~\cite{thesis}.
According to Ref~\cite{wim2ph}, one may expect that the most important
of these terms involve the couplings to two ph phonons as depicted in 
Fig.~\ref{fig:2PiERPA}.
It is not difficult to see that different diagrams, similar to those in
Fig.~\ref{fig:2PiERPA}, can be obtained through Pauli exchange of the
phonon's external lines. In total there are sixteen such possible
contributions, corresponding to all the possibilities of connecting
two ph phonons to a ph state by means of a single interaction,
both in the upper and lower part of the diagram.
 It is this approximation to the irreducible interaction that will be
pursued in the present work.

An additional ingredient entering the 2p2h Extended RPA (ERPA) of
Refs.~\cite{brand88,brand90,EDRPAgeurts} requires further discussion.
This involves diagrams similar to the one
in Fig.~\ref{fig:ERPAflip}b that are obtained from the 2p2h contribution
by inverting the sense of propagation of either the incoming or the outgoing
ph pair. These diagrams involve higher excitations (at least
3p3h, when combined in the expansion of Fig.~\ref{fig:RPA_exp}) and
are expected to give rather small contributions.
 Nevertheless, they  represent corrections to the terms of the ph interactions
that control the RPA correlations and they also add Pauli corrections
to the RPA expansion of Fig.~\ref{fig:RPA_exp} at the 3p3h level.
 In Ref.~\cite{brand90} it was found that they play a role
in  stabilizing some particular solutions.

The present calculation includes both the direct two-phonon contributions
of the type depicted in Fig.~\ref{fig:2PiERPA} and the diagrams
similar to the one in Fig.~\ref{fig:ERPAflip}c. Obviously, all the 
2p2h ERPA contributions are incorporated in this approach. 
Moreover, full two-phonon configurations are accouned for and a
SCGF approach is applied.
 Thus, the present formalism 
is an extension of (and goes well beyond) the calculations of
Refs.~\cite{brand88,brand90,EDRPAgeurts,norm1}.  However, to avoid
complications in the notation, we will still refer to
it as ``Extended RPA'', or as ``two-phonon ERPA'' whenever confusion
may arise.

The Baym-Kadanoff procedure also generates 
other two-phonon contributions, for example those coupling a
pp and a hh phonon. These could be mixed with the
presently considered configurations by means of an all order expansion of the
Faddeev-Yakubovsky type.
 Given our present knowledge, such a massive resummation of diagrams
does not appear to  be relevant for the understanding of the spectrum
of ${}^{16}\textrm{O}$.
Such a study is in any case beyond the scope of the present paper.

\subsection{Two-phonon contributions to the ph propagator}
\label{sec:ERPA_eq}

The usual DRPA equations are obtained from Eq.~(\ref{eq:BS_eq}) by choosing
$K^{(ph)}_{\alpha \beta , \gamma \delta} = V_{\alpha \delta ,\beta \gamma}$,
as mentioned above, and by keeping dressed propagators as external lines.
In this way, one is left with only two-time quantities and, after Fourier
transformation, the DRPA equation becomes 
\begin{equation}
  \Pi(\omega)_{\alpha  \beta  , \gamma  \delta} = ~
    \Pi^{f}_{\alpha  \beta  , \gamma  \delta}(\omega) ~+~ 
     \sum_{\mu , \rho , \nu , \sigma} ~
       \Pi^{f}_{\alpha  \beta  , \mu  \rho}(\omega) ~
          V_{\mu  \sigma  , \rho \nu} ~
           \Pi_{\nu  \sigma  , \gamma  \delta}(\omega) \; ,
 \label{eq:DRPA}
\end{equation}
where all the indices and summations are shown explicitly.
In Eq.~(\ref{eq:DRPA}), the {\em free} polarization propagator
$\Pi^{f}(\omega)$ is also a two-time quantity, with the
following Lehmann representation
\begin{equation}
\Pi^{f}_{\alpha \beta ,
       \gamma \delta}(\omega) = ~ ~ \sum_{n , k}  ~
 \frac{\left( {\cal X}^{n}_{\alpha}{\cal Y}^{k}_{\beta} \right)^* 
            \;{\cal X}^{n}_{\gamma}{\cal Y}^{k}_{\delta}}
     {\omega - \left( \varepsilon^{+}_{n} 
             - \varepsilon^{-}_k \right) + i \eta }
 ~+~  \sum_{k , n}    ~
 \frac{  {\cal Y}^{k}_{\alpha}{\cal X}^{n}_{\beta} \; 
    \left( {\cal Y}^{k}_{\gamma}{\cal X}^{n}_{\delta} \right)^*  }
  {\omega + \left( \varepsilon^{+}_{n} 
                 - \varepsilon^{-}_{k} \right) - i \eta } ,
\label{eq:Pif}
\end{equation}
where ${\cal X}^{n}_{\alpha} = {\mbox{$\langle {\Psi^{A+1}_n} \vert $}}
 c^{\dag}_\alpha {\mbox{$\vert {\Psi^A_0} \rangle$}}$%
~(${\cal Y}^{k}_{\alpha} = {\mbox{$\langle {\Psi^{A-1}_k} \vert $}}
 c_\alpha {\mbox{$\vert {\Psi^A_0} \rangle$}}$) are the
spectroscopic amplitudes for the excited states of a system with
$A+1$~($A-1$) particles and the associated 
poles $\varepsilon^{+}_n = E^{A+1}_n - E^A_0$%
~($\varepsilon^{-}_k = E^A_0 - E^{A-1}_k$) correspond to the excitation
energies with respect to the $A$-body ground state.
The indices $n$ and $k$ label the eigenstates of the
systems with $A \pm 1$ particles and enumerate the fragments associated
with the one-particle and one-hole excitations, respectively. 
When a dressed  propagator is used as input, its one-body overlap functions
and quasiparticle energies already contain information about the coupling of
sp motion to 2p1h, 2h1p and more complex configurations.
 As a consequence, contributions
beyond the 1p1h case are already included in~$\Pi^{f}(\omega)$.

 Methods based on an RPA-like expansion produce an infinite series of
diagrams in which the direction of propagation can be reversed from
backward to forward and vice versa. In the case of standard (D)RPA,
the interaction kernel is simply given the potential $V$ and it is the same
for every contribution to the diagrammatic expansion. Therefore the usual
(D)RPA equation can be written in compact form, as in Eq.~(\ref{eq:DRPA}).
 This is no longer true when one aims to include additional contributions and
at the same time insists on working with two-time quantities. The
two-phonon diagrams of Figs.~\ref{fig:2PiERPA} and~\ref{fig:ERPAflip}c have
different analytical expressions. As a consequence one first needs to 
separate all the four possible time directions of the kernel $K^{(ph)}$
--forward to forward, backward to backward, and the two time-inversion cases--
before including the relevant diagrams in the Bethe-Salpeter
equation~(\ref{eq:BS_eq}).
This can be achieved by splitting the free ph propagator~(\ref{eq:Pif})
into its forward- and backward-going parts, denoted by ${}^ >$ and ${}^<$
respectively,
\begin{equation}
     \Pi^f(\omega) \longrightarrow   \Pi^{f~>}(\omega)  ~+~ \Pi^{f~<}(\omega)
     \;  . 
\label{eq:split_Pif}
\end{equation}
By performing this substitution in Eq.~(\ref{eq:BS_eq}) one
obtains a similar separation for the complete ph propagator
\begin{equation}
     \Pi(\omega) \longrightarrow   \Pi^>(\omega)  ~+~ \Pi^<(\omega) \; ,
\label{eq:split_Pi}
\end{equation}
where ${}^ >$ and ${}^<$ now refer to the sense of propagation of the
final lines only. Suppressing the indices and summations one obtains
\begin{eqnarray}
   \Pi^>(\omega) &=& \Pi^{f~>}(\omega) ~+~ 
       \Pi^{f~>}(\omega) ~ K^{(ph)} ~ \Pi(\omega)  \; ,
\nonumber \\
  \Pi^<(\omega) &=& \Pi^{f~<}(\omega) ~+~ 
       \Pi^{f~<}(\omega) ~ K^{(ph)} ~ \Pi(\omega) \; .
 \label{eq:split_BS}
\end{eqnarray}
The last step consists in substituting Eq.~(\ref{eq:split_Pi})
into~(\ref{eq:split_BS}) and approximating each component of the ph kernel
$K^{(ph)}$ with the sum of the bare interaction and the corresponding
two-phonon contributions. The result corresponds to 
the ERPA equations given by
\begin{eqnarray}
  \Pi^>(\omega) &=&  \Pi^{f~>}(\omega) ~+~  
    \Pi^{f~>}(\omega) ~
       \left\{ \left(V ~+~ W^>(\omega) \right)  \Pi^>(\omega)
             + \left(V ~+~ H^{>,<}     \right)  \Pi^<(\omega) \right\}
\nonumber \\
\nonumber \\
   \Pi^<(\omega) &=& \Pi^{f~<}(\omega) ~+~ 
     \Pi^{f~<}(\omega) ~ 
       \left\{ \left(V ~+~ H^{<,>}     \right)  \Pi^>(\omega)
             + \left(V ~+~ W^<(\omega) \right)  \Pi^<(\omega) \right\} \; .
\label{eq:2PiERPA}
\end{eqnarray}
 In Eq.~(\ref{eq:2PiERPA}), $W^>(\omega)$ represents the contribution of
all the sixteen two-phonon diagrams, Fig.~\ref{fig:2PiERPA}, in the forward
direction. $W^<(\omega)$ corresponds to the contributions connecting the
backward-going terms.
 Analogously $H^{>,<}$ represent the sum of the diagrams in which a
backward-going hp state is inverted in a forward-going ph one
(Fig.~\ref{fig:ERPAflip}c illustrates such a case)  while $H^{>,<}$ is the 
time-reversed contribution.

The practical implementation of Eqs.~(\ref{eq:2PiERPA}) requires additional 
manipulation to treat the freely propagating lines in Figs.~\ref{fig:2PiERPA}
and~\ref{fig:ERPAflip}c. This situation is completely analogous to the
one already discussed in Ref.~\cite{FaddRPA1,BaDLund} 
for the 2p1h expansion of the self-energy.
Specific details for the ERPA equation~(\ref{eq:2PiERPA}) are
given in the sections in the Appendix.
The next section reports on the
application of this formalism to ${}^{16}\textrm{O}$.

\section{Results}
\label{sec:results}

As in the work of Ref.~\cite{FaddRPAO16}, the ERPA equations where
solved within a model space consisting of a finite set of 
harmonic oscillator states. All the first four major
shells (from $1s$ to $2p1f$) plus the $1g_{9/2}$ where included to account
for the sp orbitals that are most relevant for low-lying excitations.
The harmonic oscillator parameter was chosen to be $b$=1.76~fm. 
As a consequence of the truncation of the model space a Brueckner $G$-matrix
was used as a microscopic effective interaction.  
This $G$-matrix
was derived from a Bonn-C potential~\cite{bonnc} and computed according
to Ref.~\cite{CALGM}. 

 The contributions of two-phonon states were first studied by solving the
ERPA equation~(\ref{eq:2PiERPA}) with an independent-particle model (IPM)
propagator.
This propagator was constructed
from a Slater determinant composed of the lowest occupied 
harmonic oscillator (h.o.) orbitals in the model space.
Where the effects  of nuclear fragmentation were included,
the calculations employed the fully dressed sp propagator 
computed in Ref.~\cite{FaddRPAO16}.
 This propagator contains no more than two main quasihole (quasiparticle)
fragments for each state in the $p$ ($sd$) shell. For these orbits, the
sp propagator can therefore be well represented by means of one or two
principal fragments and an effective pole that accounts for strength
far from the Fermi energy.
 For quasiparticle states associated with the $pf$ shell, instead, the
fragmentation is more substantial~\cite{thesis} and the self-consistent
propagator  was approximated by including the most important poles
(up to 4 main fragments for the $p_{3/2}$) plus two effective ones that
gather all the background strength.
 We have checked that the results are not sensitive to the details of the
fragmentation of the $1f$ orbitals. 
For the $2p$ orbitals, we found that the strength distribution is
sufficiently represented by the adopted prescription and further discuss
the sensitivity to the results in Sec.~\ref{sec:instab_0+}.

In both cases (IMP and dressed inputs), the standard (D)RPA equation was
solved first, in order
to generate the ph phonons that entered the final two-phonon ERPA calculation.
 When coupling different phonons,
only the lowest few states of ${}^{16}\textrm{O}$ have the right quantum
numbers, and energies low enough, to generate two-phonon
contributions with unperturbed energy below 20~MeV.
These are the only states  relevant for our purposes~\cite{wim2ph}.
 In practice, the calculations were performed by including, in each channel,
all two-phonon configurations up to 30~MeV.
 The stability tests reported in Sec.~\ref{sec:furt_stab} 
demonstrate that, for the case of a dressed input propagator, 
higher two-phonon contributions do not change the results
for the low-lying states appreciably.

\subsection{Results for the particle-hole propagator}

\begin{figure}
 \begin{center}
 \includegraphics[width=4.5in]{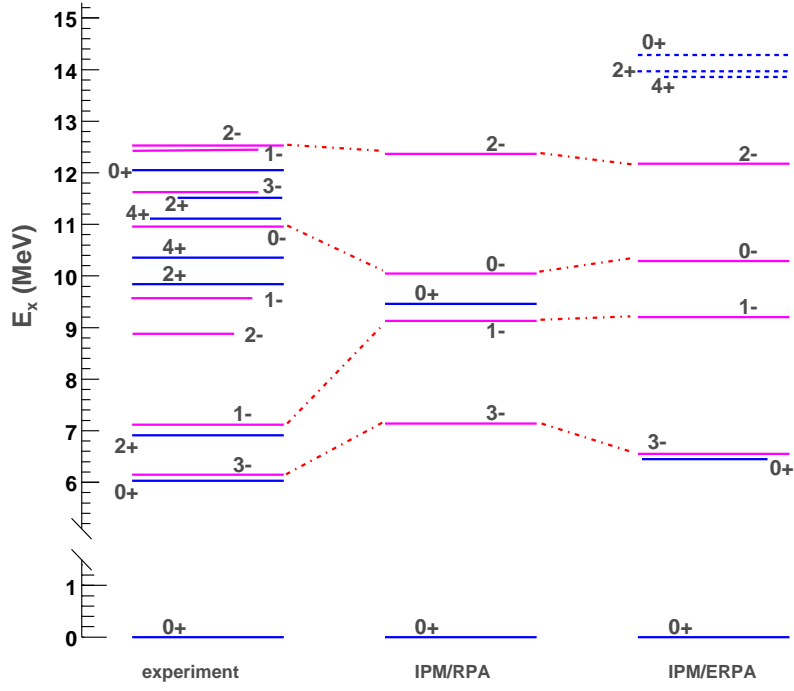}
 \end{center}
\caption[Two-phonon ERPA results for ${}^{16}\textrm{O}$; h.o. IPM input]
 {\small
 Results for the two-phonon ERPA propagator of ${}^{16}\textrm{O}$ with an 
 h.o. IPM input propagator, last column.
  The spectrum in the middle is obtained by solving the standard RPA
 problem and is employed, as is, to generate two-phonon contributions
 to the ERPA equation.  
  The excited states indicated by dashed lines are those for which the ERPA
 equation predicts a total spectral strength $Z_{n_\pi}$ lower than 10\%.
  The first column reports the experimental results taken from Ref.~\cite{azj}.
  \label{fig:2PiERPA_ho} }
\end{figure}

The results obtained from the h.o. input propagator 
are displayed in Fig.~\ref{fig:2PiERPA_ho}.  The isoscalar 
eigenvalues obtained for energies below 15~MeV are displayed, for both
the standard RPA and for the ERPA calculation.
The ERPA calculation produces lower energies for some of the states that
were already obtained in RPA. Both $3^-$ and $0^+$  shifted down to
about 0.5~MeV above the experimental energy.
 As discussed further in Sec.~\ref{sec:furt_stab}, this is to be considered
rather fortuitous and one must remember that the dressing of the sp propagator,
which has the effect to raise these eigenvalues, has not been taken
into account yet.
Still, this result indicates
that correlations between ph and two-phonon states can be substantial
and go in the right direction in explaining the experimental results.

It is also worth considering the total ph spectral strength of these states
obtained by summing the corresponding amplitudes,
Eq.~(\ref{eq:PiAmpl_def}), as follows
\begin{equation}
 Z_{n_\pi} ~=~  \sum_{\alpha \beta} ~
      \left| {\cal Z}^{n_\pi}_{\alpha \beta} \right|^2   \; .
\label{eq:Z_tot_ph}
\end{equation}
Results for both the excitation energy and $Z_{n_\pi}$ of the principal levels
in Fig.~\ref{fig:2PiERPA_ho} are given in Table~\ref{tab:2PiERPA_ho}.
For the ERPA case, also the fraction of quasiparticle-quasihole and
two-phonon configuration that appear in the wave function are shown.
 Note that $Z_{n_\pi}$ is substantially bigger than one for the $3^-$
and $0^+$ states and that these values increase further for the ERPA results.
This signals an increase of the collective character of these solutions which
may lead to an instability of the RPA equations for interactions that are even 
more attractive.
The two-phonon ERPA approach generates a  triplet of states at 
about 14~MeV, with quantum numbers $0^+$,  $2^+$ and $4^+$.
 However, these levels are almost exclusively composed of
two-phonon contributions and contain only small admixtures
of ph states, resulting in a small ph spectral strength~ $Z_{n_\pi}$. 
The quantum numbers and energies of these states indeed
correspond closely to those obtained
by coupling two $3^-$ RPA phonons, each at 7.14~MeV.
We note that a similar triplet is found experimentally at 12.05, 11.52
and 11.10~MeV, which also corresponds to
twice the experimental energy of the first $3^-$ phonon.
The first experimental  $2^+$  is found at lower energy and 
its spectral strength  is known to have relevant ph
components~\cite{O16ph86}.  Thus it cannot be identified with any of the
above two-phonon contribution.
For all the lowest states that are not already reproduced by standard RPA
a very small ph spectral strength has been found, due to a general lack of
mixing between the ph and the two-phonon configurations.
 Of interest is also the $2^+$ that represents the giant quadrupole resonance
at 20.7~MeV. In this case RPA and ERPA give 22.9 and 23.3~MeV for the main
peak but
with a lower $Z_{n_\pi}$ in the second case. For this state,
part of the ph strength (about 20\%) is shifted to two-phonon configurations
representing the expected fragmentation of the giant resonance.

\begin{figure}
 \begin{center}
 \includegraphics[width=4.5in]{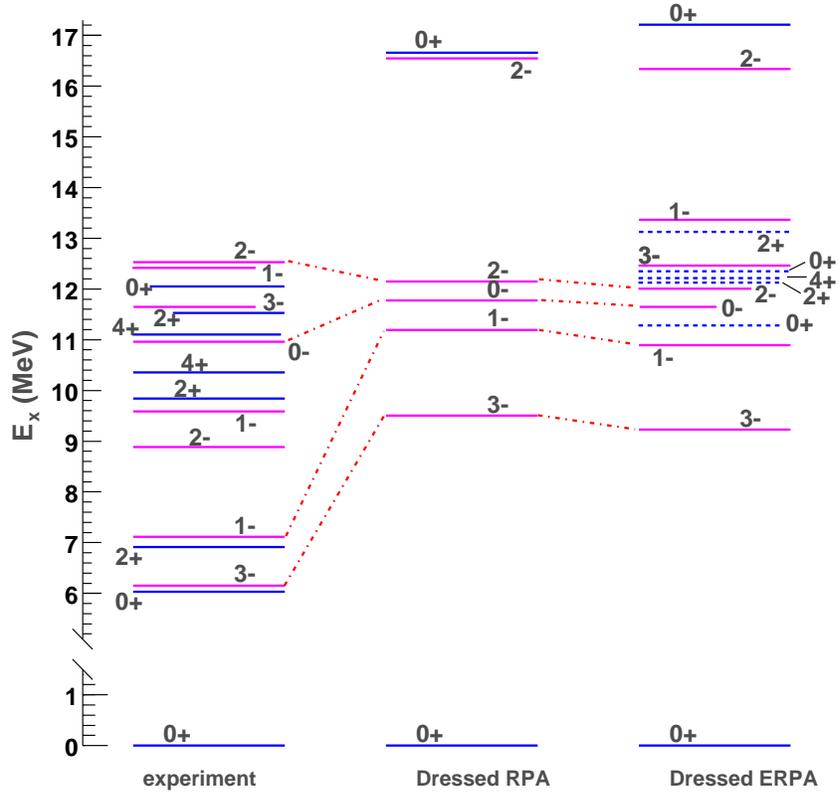}
 \end{center}
\caption[Two-phonon ERPA results for ${}^{16}\textrm{O}$; dressed input]
 {\small 
 Results for the DRPA and the two-phonon ERPA propagator of ${}^{16}\textrm{O}$
 with a dressed input propagator from Ref~\cite{FaddRPAO16}, middle and
 last column respectively.
  In solving the ERPA equation, the lowest $3^-$, $1^-$ and $0^+$ levels
 of the DRPA propagator where shifted to their experimental energies. All
 other DRPA solutions were left unchanged.
 The excited states indicated by dashed lines are those for which the (E)RPA
 equation predicts a total spectral strength $Z_{n_\pi}$ lower than 10\%.
  The first column reports the experimental results~\cite{azj}.
\label{fig:2PiERPA_3g} }
\end{figure}

Fig.~\ref{fig:2PiERPA_3g} and Table~\ref{tab:2PiERPA_3g} show  the analogous
results when dressed sp propagators are used as input. In general, the main
effect of fragmentation is to screen the nuclear interaction as a consequence
of the quenching of spectroscopic factors for the input sp propagator.
 For the DRPA case, this results in  increasing the lowest $3^-$ and $1^-$
solutions by $\sim$2~MeV.
More substantial is the effect on the lowest $0^+$ state with a predominantly
ph character which rises to
about 17~MeV, confirming the sensitivity of this state to details
of the fragmentation and the strength of the nuclear interaction.
Unlike the IPM case, we have chosen to solve the two-phonon ERPA by 
first shifting
the lowest solution for both the $0^+$, $3^-$ and $1^-$ states down to their
relative experimental energies.
This has the advantage of lowering the most important two-phonon
configurations and allows to investigate their interplay with the ph ones.

As can be seen in Fig.~\ref{fig:2PiERPA_3g}, the ERPA equations still
generate a triplet of $0^+$, $2^+$ and $4^+$ states at twice the
energy of the first $3^-$ phonon. Due to the screening of the ph interaction,
these states mix very little with the ph configurations yielding
an almost degenerate triplet. Table~\ref{tab:2PiERPA_3g} also gives a
comparison between the total two-phonon content of the states and the
individual contributions of the
most important configurations.
This decomposition demonstrates that this triplet
is formed by pure $3^- \otimes 3^-$ states.
 This observation is in accordance with the 2p2h character of these
states~\cite{O16ph86,O162p2h}.
  Since the overall energy of
these states approximately agrees with experiment, one can expect that
this calculation correctly represents the bulk properties of their 
wavefunctions.
Nevertheless,
it is clear that an additional interaction is needed in order to split
this triplet as observed experimentally.
 It should be noted that no solution that can be identified with
the $2^+$ and $4^+$ levels at 9.8 and 10.4~MeV. The $2^+$ level has been
interpreted in terms of an $\alpha$ particle rotating around an
excited ${}^{12}\textrm{C}$ core~\cite{O16ph86} and therefore involves
correlations that may go well beyond the present calculation.
With regard to the $2^+$ strength around 20~MeV, we note that the sp
fragmentation included in the DRPA equation already generates a distribution
of $2^+$ strength. Table~\ref{tab:2PiERPA_3g} contains a few of these
solutions and shows that also for the dressed case a sizable mixing of
ph and two-phonon
contributions is obtained at this energy. This mixing generates the 
spreading of ph strength over different solutions, consistent with
the finite width of such a resonant state.
 It should be noted, however, that a discrete basis is used here and therefore
it is not possible to properly describe the continuous strength
distribution of a giant resonance, for
which a continuum or a complex basis should be used~\cite{berggren}.

 Also interesting are the results for the other isoscalar
$0^+$ states. The lowest
solution with a predominant ph character, that was found in DRPA
at $\sim$17~MeV,
is obtained at a similar energy but is now characterized by a partial
mixing with two-phonon configurations. The ph and two-phonon contents of
this state given in Table~\ref{tab:2PiERPA_3g} show that this is the result
of mixing with the 
the lowest solution, which ends up at $\sim$11~MeV. The latter is
predominately a two-phonon state.  It is also seen that in both cases the
relevant configuration comes from the coupling of two $0^+$ phonons.
Configurations involving two $3^-$'s make smaller contributions
while the other modes not reported in the table were negligible.
 We note that the wavefunctions for these
states contain several relevant ph configurations, obtained
from different quasiparticle fragments in the $pf$~($sd$) shells combined 
with quasihole fragments of the $p$~($s$) shells. 
 Therefore, the situation is more complicated than the simple picture
of only two levels interacting with each other.
We observe that the shell-model calculations for the first excited 
state of ${}^{16}\textrm{O}$ give very small contributions from both
0$\hbar\omega$ and 2$\hbar\omega$ configurations and a strong
population of 4$\hbar\omega$ states~\cite{Haxton0+,wbm92}. On the other
hand, inelastic electron scattering experiments~\cite{O16ph86}
clearly excite this
state identifying its partial ph character even though the state is dominated
by 4p4h components.
 From the point of view of the 
SCGF approach, the one-body response is completely
described by the (dressed) ph propagator, Eq.~(\ref{eq:DRPA}), and therefore
the total one-body strength must be represented
by $Z_{n_\pi}$~(\ref{eq:Z_tot_ph}).
 This observation
points to the necessity of a stronger mixing between ph and two-phonon
configurations than obtained in the present calculations.
 Still, the present results indicate that the two-phonon configuration
obtained by
coupling the two lowest $0^+$ phonons plays a role in
the structure of the first $0^+$ state itself. This means that, starting
from the dressed ph admixture contained in it, a fully self-consistent
calculation would also generate some contributions of 2p2h, 4p4h, and beyond.
 The calculations of Ref.~\cite{iach1} have shown that the bulk of
the 4p4h contributions to the first excited state of ${}^{16}\textrm{O}$
may come from the coupling of four different phonons with negative
parity ($3^-$ and $1^-$). However,
the self-consistent role of  coupling positive parity states was
not conidered in that work. Both this effect,
the RPA correlations and the inclusion of the nuclear fragmentation allow
for the --at least partial-- inclusion of configurations beyond the 2p2h case
even when no more than two-phonon coupling is considered, as in this paper.
 A study of the  importance of three- and four-phonon configurations
within this approach is beyond the scope of the present work.

 The principal negative parity isoscalar states are still shifted down in 
energy by ERPA over DRPA, as expected, but the improvement is less
than 0.5~MeV. 
The low-lying $3^-$ and $1^-$ levels remain substantially
above the experimental energy at 9.23 and 10.90~MeV, respectively,
and more correlations will be needed in order to lower their energy.
We note that these wavefunctions contain a two-phonon admixture obtained by
coupling the low-lying $0^+$ excitation to either the $3^-$ or $1^-$ phonons.
 According to the above interpretation of the first $0^+$ excited state
this corresponds to the inclusion of 3p3h and beyond.
 At the same time two additional negative parity solutions, with two-phonon
character, are found at higher energy in accordance with experiment.

The  ERPA results for an h.o. input  give appropriate corrections to
several energy levels.
 However, such corrections tend to become negligible in the successive
calculations employing dressed propagators.
 Presumably this happens because a sizable part of the correlations, which in 
the latter case are introduced by ERPA, are already accounted for by
the dressing of the sp propagators. 
In general, the inclusion of fragmentation has also the effect of screening the 
interaction between ph and two-phonon configurations. However, the results
of Table~\ref{tab:2PiERPA_3g} show that a considerable mixing between the 
two can be obtained if all the relevant configurations are sufficiently
low in energy, although no substantial mixing between different 
two-phonons states is seen.  According to these considerations, the biggest
deficiency of the present approach is probably the lack of an interaction 
within the multi-phonon space.  Refs.~\cite{iach1,iach2} have shown that
such correlations are due to pp and hh interactions, neglected here,
and that their
inclusion has important effects on the final result for the spectrum.
 These terms may considerably 
lower most of the eigenstates of the ERPA equations.
 However, a second type of correlation between multi-phonon states, that
was not accounted for in Refs.~\cite{iach1,iach2}, is the Pauli exchange 
between the fermion lines included in different phonon propagators. These
will probably generate the opposite effect, screening part of the pp
and hh correlations.

Although the results obtained with a dressed input propagator
for the lowest excited states are not completely successful,
 it must be
recognized that the present ERPA calculation does represent a step forward.
Indeed a proper description of two-phonon contributions,
at the present level, had not been accomplished before.
 These states not only account for
configurations at the 2p2h and higher level but appear to have some relevance
for the description of states at higher energy.
In addition, excitations are generated with
quantum numbers that can not be obtained in the standard (D)RPA approach
at these low energies.
Given the results of the present work, it appears  that a 
calculation including both pp and hh correlations, a more complete treatment
of Pauli effects and the most relevant configurations up to the
four-phonon states can be achieved with the presnt-day computers.
Such a calculation is planned for the future.
 
\section{Stability of the results and truncation of the model space}
\label{sec:trunc}

\subsection{RPA instability}
\label{sec:instab_0+}

 The RPA approach may break down for strongly attractive 
interactions, by genereating
an excess of collectivity and lowering the first excited state
below the ground-state energy.
 The inclusion of fragmentation of the sp strength
can affect this behavior in several ways.
First, both the screening of the nuclear interaction and the splitting of
the particle and hole spectral strengths over different fragments
tends to stabilize the DRPA equations.
Second, correlations lower most of the sp energies in
comparison with the IPM. This results in smaller unperturbed ph energies and
therefore has the opposite effect of pushing the DRPA approach toward
instability.
 Third, the instability can be sensitive to the details of the spectroscopic
amplitudes.
 This last point can be best illustrated by looking at the sp strength
with $j^{\pi}=3/2^-$, to which the present calculation is most sensitive.
For the model space employed in this work a given particle overlap function
is a superposition of two h.o. eigenstates,
\begin{equation}
\psi^{n}_{p_{3/2}}({\bf x}) ~=~ 
       {\cal X}^{n}_{1p_{3/2}}  \; \phi^{h.o.}_{1p_{3/2}}({\bf x})  ~\pm~
       {\cal X}^{n}_{2p_{3/2}}  \; \phi^{h.o.}_{2p_{3/2}}({\bf x}) 
     \; ,
\label{eq:overlap}
\end{equation}
 where we have stressed the fact the two components can sum either
constructively or destructively. The fragmentation introduces a
non-zero ${\cal X}^{n}_{1p_{3/2}}$ component even for particle
states that in the IPM would be described only by the $2p_{3/2}$ subshell.
By considering Eqs.~(\ref{eq:DRPA-Bmtx}) one can see that
the ph interaction for the DRPA approach can be substantially changed by this
new component by
virtue of the strong matrix elements of the effective interaction between 
$1p_{3/2}$ states. In particular, this also applies to the off-diagonal terms
that drive the RPA-like correlations. It is also worth nothing that the sign
of this correction changes accordingly to the relative sign of the
components in Eq.~(\ref{eq:overlap}), therefore pushing the states 
either towards stability or instability, respectively.

We have observed that choosing a wrong sign for one or more of the $p_{3/2}$ 
particle fragments has drastic effects for the solutions of the isoscalar
$0^+$ channel, with the choice of constructive interference in
Eq.~(\ref{eq:overlap}) leading toward instability of the DRPA approach.
 For the input propagator employed in this work, all of the background
fragments have a similar wavefunction and a negative or positive
interference sign according to whether their sp energy is
lower or higher than 20~MeV, respectively~\cite{thesis}.
 Therefore it is natural to collect such background
distributions in two different poles, one for each region of missing energy.
 A similar behavior applies also to the other quasiparticle states in the
$2p$ and $sd$ shells.
We have checked that the results of Sec.~\ref{sec:results} are
stable with respect to the number of main fragments and effective poles
included in each subshell, therefore no instability affects the
present work.
 However, in applications in which the number of poles has to be strongly
reduced it may not be possible to chose effective poles of the sp propagator
that represent the details of the spectral distribution sufficiently
accurately.
 In such a case an instability in the DRPA approach may arise.
 This situation occurred with
a previous DRPA calculation of the isoscalar
$0^+$ channel, reported in Ref.~\cite{FaddRPAO16},
in which a smaller number of fragments was included.  Thanks to improvements 
in the computer code that solves the DRPA problem, it is now possible
to consider a sufficiently large number of poles in the sp propagator and
this problem has been overcome.

\subsection{Time-inversion diagrams}

The contribution of two-phonon terms to the time inversion diagrams of 
Fig.~\ref{fig:ERPAflip} can, in  principle, generate other correction to the
interaction between ph and hp states that drives the RPA correlations.
Nevertheless, the effect is very small due to the large energy denominators
that appear in Eqs.~(\ref{ERPA-Bmtx1}) and~(\ref{ERPA-Bmtx2}).
We have tested their influence by neglecting the corresponding
terms $H^{>,<}$ and~$H^{<,>}$ in Eq.~(\ref{eq:2PiERPA}) and report
the results in Table~\ref{tab:B_mtx}. No appreciable change is
generated illustrated by 
differences with respect to the results of Sec.~\ref{sec:results} of  
at most 0.1\%.
 It is worth noting that the contribution of $H^{>,<}$ and~$H^{<,>}$ 
can in principle carry information on the Pauli breaking at the
level of 3p3h and beyond. Therefore, they may become more important for
the case of three-phonon calculations.  Nevertheless, Table~\ref{tab:B_mtx}
suggest that they are not likely to play an important role in the
description of the spectrum of~${}^{16}\textrm{O}$.

\subsection{Stability vs. number of two-phonon configurations}
\label{sec:furt_stab}

\begin{figure}
 \begin{center}
 \includegraphics[width=3in]{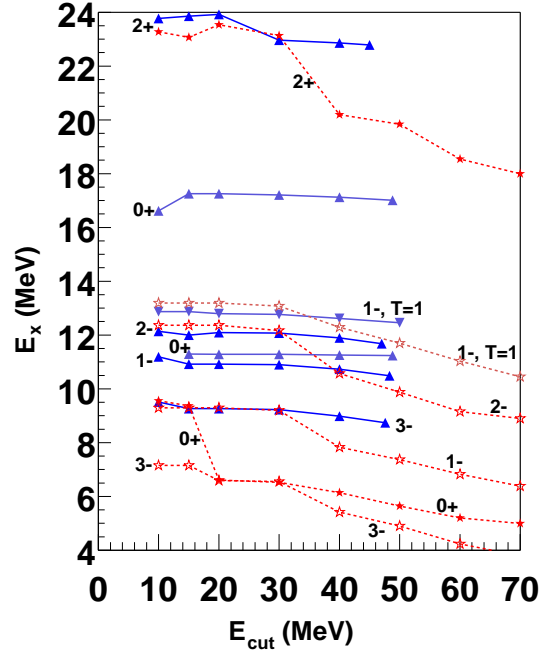}
 \end{center}
\caption[]{Dependence of the ERPA solutions on the number of two-phonon
   states considered.  For any given point, all the configuration with
   energy  $\varepsilon^\pi_{n_a} + \varepsilon^\pi_{n_b} \leq E_{cut}$
   have been included in the calculation.
    Solid~(dashed) lines refer to results obtained from a dressed~(IPM)
   input propagator.
\label{fig:2phon_trunc} }
\end{figure}

Fig.~\ref{fig:2phon_trunc} shows the results for selected solutions
of the ERPA equations, obtained by employing different sets of two-phonon
states.
 For any given point, only those configuration with energy
$\varepsilon^\pi_{n_a} + \varepsilon^\pi_{n_b} \leq E_{cut}$ have been
included in the calculation.
For $E_{cut}$=70~MeV, about 700 to 1000 two-phonon contributions have been
included, depending on the channel. This is one order of magnitude
larger than the number of ph states that enter the calculation.
 Since no two-phonon configuration has energy lower than $10$~MeV, the
leftmost points in Fig.~\ref{fig:2phon_trunc} correspond to the simple
(D)RPA calculation.
 As the lowest few two-phonon contributions are included, the solutions for
these levels show a sudden jump for all the lowest excited states.
 After this, the results obtained by using an IPM input still continue to
exhibit a 
dependence on the value of the cut-off, roughly lowering states
by $1$~MeV every time 
that $E_{cut}$ increases by $20$~MeV.
In particular, the $0^+$ and $3^-$ solutions decrease progressively,
eventually heading to RPA instability. This confirms that the 
values reported in Table~\ref{tab:2PiERPA_ho} for these states 
do not correspond to converged results.
The situation is much better for the case of a  dressed input
propagator, for which the low-lying solutions
are approximately stable for $E_{cut} >$~20~MeV.
This confirms that corresponding higher-energy two-phonon
excitations do not have strong influence
as their effects have already been included by the dressing of the
sp motion.

In Fig.~\ref{fig:2phon_trunc} we also show the results for the most
relevant of the $2^+$ solutions
that represent the resonance at about $20$~MeV. Here the variation is more
appreciable, in particular for values of the cut-off $E_{cut}$ comparable
with the energy of the state itself.
Also here a more stable behavior is obtained for the case of a dressed input
propagator.
 We conclude that when the nuclear fragmentation is accounted for the choice
of $E_{cut}=30$~MeV  adopted in Sec.~\ref{sec:results} is a adequate
for considering the low-lying spectrum.

\subsection{Size of the model space}

As a second possible source of uncertainty one may consider the size
of the employed model space. The choice of the model space employed
in Sec.~\ref{sec:results} allows to
take into account excitations up to the $2p1f$ shells. This means
that ph configurations up to $2\hbar\omega \sim 27$~MeV are included,
consistently with the choice for truncating two-phonon states discussed above.
The calculations of Ref.~\cite{brand88,EDRPAgeurts} also show that
this is adequate and suggest that
better convergence is to be expected for a calculation with a dressed input.
To check this for the present calculation, we have augmented
the model space by adding two more h.o. shells, up to $\rho = 2n+\ell$=5,
and recomputed the $G$-matrix elements accordingly.
 The input sp propagator used in the calculations was obtained by including
an extra particle pole for every subshell added to both the IPM's Slater
determinant and the fragmented dressed propagator.  
The sp energies for the new shells were chosen by solving the
Brueckner-Hartree-Fock equation within the new model space, in an approach
analogous to the one of Ref.~\cite{GeurtsO16}.

The results are compared to the ones obtained for the smaller space in
Table~\ref{tab:B_mtx}.
For the IPM, a sizable variation of the $3^-$ excitation energy is found.
Also, the high values of the ph strength $Z_{n_\pi}$ for both the $0^+$
and $3^-$ solutions indicate that these states are approaching instability.
 As a consequence of the different $3^-$ excitation energy,
which also enters the two-phonon calculation, most of the ERPA solutions are 
 shifted down in energy. The main conclusions of
Sec.~\ref{sec:results}, however, remain unchanged.
When dressed sp propagators are employed, no dramatic change occurs for
the low-lying solutions and only some states with dominant ph character are
shifted by no more that 1~MeV. 
 This suggests that the main conclusions of
Sec.~\ref{sec:results} are also not affected by the truncation of the model
space, whence the bulk of the missing correlations has to be looked for
in an extension of the diagrammatic expansion.
We note, however, that a more attractive G-matrix is
expected to be obtained if the continuum outside the model
space were modified by accounting for self-energy corrections.
 This feature may lead to a further improvement
of the present results.
In addition, one may explore the inclusion of effective three-body
force as due to propagation outside the model space.
Such effective forces are critical in obtaining the correct binding energy
in no-core shell model calculations reported in Ref.~\cite{3Nforce}.

\section{Conclusions}
\label{sec:concl}

 The dressed RPA equations have been extended to account for the coupling
of two ph phonons in forming the excited states of a many-body system.
The coupling among ph and multi-phonon configurations is conceptually
similar to the Interacting Boson Model of Ref.~\cite{iach1} but limited, in
this application, to only two-phonon admixtures.  Nevertheless, the present
approach has the added advantage of taking into account both the effects
of nuclear fragmentation and the RPA-like correlations, as generated by
two-phonon fluctuations in the ground state.

The resulting Extended RPA formalism has been applied to study the excitation
spectrum of ${}^{16}\textrm{O}$.
The results suggests a sizable mixing of ph and two-phonon configurations
for the case of the $2^+$ quadrupole resonance at 20.7~MeV.
Other solutions, carrying quantum numbers that cannot be generated by the
simple DRPA, are obtained with this method,
among which a few isoscalar negative parity states and a triplet
with $J^\pi$=$0^+$, $2^+$ and $4^+$ near 12~MeV,
in accord with experiment.
 In particular, the states in the triplet were seen to be almost pure
two-phonon configurations obtained by coupling two 
$3^-$ phonons. This feature confirms the 2p2h character of these states.
The results are less satisfying for the low-lying positive parity states,
which are known to require a proper description of 4p4h excitations.
The present approach predicts the lowest solution for the isoscalar $0^+$
channel to be at 11.3~MeV, considerably above the first experimental
excited state. A sizable component of this solution is seen to be generated
by the coupling of two of the lowest  $0^+$ phonon themselves. This suggests
that important contribution of 4p4h excitation may be included already at the 
two-phonon level in a self-consistent fashion.
 Other correlations beyond the 2p2h level are partially included
in the present calculation through the dressing of the propagators
and the RPA approach.
As expected, though, the present implementation
is not sufficient to obtain a complete
description of the low-lying excitation spectrum of ${}^{16}\textrm{O}$.
 The results obtained in this work show that an interaction between 
multi-phonon configuration is still missing. This can be achieved by 
including pp and hh correlation between different ph phonons.
Also the inclusion of up to four-phonon states is expected to be relevant
for this system.

Finally, the stability of the present results has been tested with respect
to the truncation of the model space and to the number of two-phonon
configurations accounted for. It was found that the effects of nuclear
fragmentation acts to `renormalize' the sp propagator, making the
solutions of the ERPA fairly independent of higher energy
configurations. As discussed in Sec.~\ref{sec:trunc}, this 
feature generates stable solutions with respect to
the size of the model space and the corresponding $G$-matrix
interaction used in this work.

The four-phonon calculation of Refs.~\cite{iach1,iach2} 
give a very good description of the excitation
spectrum of ${}^{16}\textrm{O}$. Those findings and the
effects of fragmentation discussed in this paper suggest that the present
ERPA formalism can be suitably extended to generate a satisfactory
description of this nucleus, within the framework of SCGF.
Such extensions are presently under consideration.
 Such a calculation  may also significantly reduce the discrepancy
between the measured and the theoretical sp spectral function for this
nucleus~\cite{FaddRPAO16}.

\acknowledgments
This work was supported in part by the U.S. National Science Foundation
under Grants No.~PHY-9900713 and PHY-0140316 and in part by the Natural
Sciences and Engineering Research Council of Canada (NSERC).

\appendix

\section{Extended DRPA Equation with Two-phonon Contributions}
\label{app:A.ERPA}

 When reducing the kernel of the Bethe-Salpeter Eq.~(\ref{eq:BS_eq}) to a 
two-time quantity, one has to deal with the fact that some lines in the
diagrammatic expansion continue to propagate unperturbed while some
interaction occurs between other particles. This is also the case for the
diagrams of Figs.~\ref{fig:2PiERPA} and~\ref{fig:ERPAflip}. This situation can
be overcome by redefining the objects that appear in Eq.~(\ref{eq:2PiERPA})
and promoting the quantum numbers $\{ n , k\}$,
--that label particle and hole fragments-- to external indices. The usual
form of the ph propagator, Eq.~(\ref{eq:Pi}), is obtained only as a last
step of the calculation by performing the sum over the $\{ n , k\}$
indices~\cite{EDRPAgeurts,FaddRPA1}.

 The separation of the propagators~$\Pi^f(\omega)$ and~$\Pi(\omega)$
into forward and backward components, outlined in Sec.~\ref{sec:ERPA_eq},
is a natural consequence of adopting this prescription.
 Indeed particle and hole external lines turn into each other by time
inversion and become quantities that depend on different
quantum numbers, i.e. the fragmentation indices $\{n,k\}$.
For this reason, Eqs.~(\ref{eq:split_Pif})
and~(\ref{eq:split_Pi}) are only formal relations and the arrow can be
substituted by an equal sign only before summing over all the
particle and hole fragments.
The free polarization propagator~(\ref{eq:Pif}) naturally splits in two
components that are purely forward and backward-going. The relevant Lehmann
representations are
\begin{eqnarray}
\Pi^{f~>}_{\alpha  n_\alpha \beta  k_\beta ,
       \gamma  n_\gamma \delta k_\delta}(\omega) &=& ~ ~
\delta_{n_\alpha , n_\gamma}  \delta_{k_\beta , k_\delta} ~
 \frac{\left( {\cal X}^{n_\alpha}_{\alpha}{\cal Y}^{k_\beta}_{\beta} \right)^* 
            \;{\cal X}^{n_\alpha}_{\gamma}{\cal Y}^{k_\beta}_{\delta}}
     {\omega - \left( \varepsilon^{+}_{n_\alpha} 
             - \varepsilon^{-}_{k_\beta} \right) + i \eta }
  \nonumber  \\
 &\equiv& G^\dag ~ \frac{1}{\omega - D + i \eta } ~ G   \; \; ,
\label{eq:Pif>}
\\
  \nonumber  \\
 \Pi^{f~<}_{\alpha  k_\alpha \beta  n_\beta ,
       \gamma  k_\gamma \delta n_\delta}(\omega) &=& ~ -
     \delta_{k_\alpha , k_\gamma}  \delta_{n_\beta , n_\delta} ~
     \frac{  {\cal Y}^{k_\alpha}_{\alpha}{\cal X}^{n_\beta}_{\beta} \; 
    \left( {\cal Y}^{k_\alpha}_{\gamma}{\cal X}^{n_\beta}_{\delta} \right)^*  }
  {\omega + \left( \varepsilon^{+}_{n_\beta} 
                 - \varepsilon^{-}_{k_\alpha} \right) - i \eta } 
  \nonumber  \\
 &\equiv& (G^*)^\dag ~ \frac{-1}{\omega + D - i \eta } ~ G^*  \; \; ,
\label{eq:Pif<}
\end{eqnarray}
where no summation is implied and a shorter notation for the unperturbed
poles and residues has been introduced. The quantities $D$ and $G$ can
be thought as matrices whose elements contain all the unperturbed poles
and residues, respectively. Note that in this case $D$ is diagonal and
depends on the fragmentation indices $\{ n , k \}$ only, while $G$ is
rectangular because its column indices depend on the model space orbitals
$\{ \alpha \}$ as well.

The separation of the complete propagator is a little more complicated.
The splitting of Eqs.~(\ref{eq:split_Pi}) and~(\ref{eq:split_BS})
involves the time direction of the outgoing lines only.
Since the RPA series contains contributions that can invert several times
the sense of propagation of the ph diagrams, both forward and backward
poles can appear in each component
\begin{eqnarray}
\Pi^{>}_{\alpha  n_\alpha \beta  k_\beta ,
       \gamma  \delta}(\omega) ~=~
 \sum_{n \neq 0} 
 \frac{\left( {\cal Z}^{>~n}_{\alpha n_\alpha \beta k_\beta} \right)^* 
           \; {\cal Z}^{n}_{\gamma \delta}}
     {\omega - \varepsilon^{\pi}_n  + i \eta }
 ~-~ \sum_{n \neq 0} 
  \frac{  {\cal Z}^{<~n}_{\beta k_\beta \alpha n_\alpha } 
           \;  \left( {\cal Z}^{n}_{\delta \gamma} \right)^* }
     {\omega + \varepsilon^{\pi}_n  - i \eta }  \; ,
\label{eq:Pi>}
 \\
\Pi^{<}_{\alpha  k_\alpha \beta  n_\beta ,
       \gamma  \delta}(\omega) ~=~
 \sum_{n \neq 0} 
 \frac{\left(  {\cal Z}^{<~n}_{ \alpha k_\alpha \beta n_\beta} \right)^* 
           \; {\cal Z}^{n}_{\gamma \delta}}
     {\omega - \varepsilon^{\pi}_n  + i \eta }
 ~-~ \sum_{n \neq 0} 
  \frac{ {\cal Z}^{>~n}_{\beta n_\beta \alpha k_\alpha} 
           \; \left( {\cal Z}^{n}_{\delta \gamma} \right)^* }
     {\omega + \varepsilon^{\pi}_n  - i \eta }  \; .
\label{eq:Pi<}
\end{eqnarray}
 In Eqs.~(\ref{eq:Pi>}) and~(\ref{eq:Pi<}) the spectroscopic amplitude
splits in two contributions ${\cal Z}^{>}$ and ${\cal Z}^{<}$.
These appears unchanged in both equations
due to the time-inversion symmetries obeyed by the ph states.
 In terms of these definitions, and applying the summation prescription,
Eq.~(\ref{eq:split_Pi}) can be exactly formulated as follow
\begin{eqnarray}
  {\cal Z}^{n}_{\alpha \beta} &=&
       \sum_{n_1 , k_2}  \left[ {\cal Z}^{>~n}_{\alpha n_1 \beta k_2} +
                    {\cal Z}^{<~n}_{\alpha k_2 \beta n_1}  \right] \; ,
\label{eq:Zf+b}
\\
   \Pi^{>}_{\alpha  \beta , \gamma  \delta}(\omega)  &=&
   \sum_{n_1 , k_2}   \left[
     \Pi^{>}_{\alpha  n_1 \beta  k_2 , \gamma  \delta}(\omega)  +
\Pi^{<}_{\alpha  k_2 \beta  n_1 , \gamma  \delta}(\omega) \right] \; .
\label{eq:Pif+b}
\end{eqnarray}

Finally, the contributions of two-phonon diagrams in the forward and backward 
direction can be expressed in Lehmann representation as well
\begin{eqnarray}
 W^{>}_{\alpha  n_\alpha \beta  k_\beta ,
       \gamma  n_\gamma \delta k_\delta}(\omega) &=&
\sum_{n_a , n_b} ~
 \frac{\left( K^{> ~ n_a  n_b}_{\alpha  n_\alpha \beta  k_\beta} \right)^* 
            \;K^{> ~ n_a  n_b}_{\gamma  n_\gamma \delta k_\delta}}
     {\omega - \left( \varepsilon^{\pi}_{n_a}  +
             \varepsilon^{\pi}_{n_b} \right) + i \eta }
\nonumber \\
     &=&  {K^>}^\dag \frac{1}{\omega -  E + i \eta} {K^>}  \; ,
\label{eq:W>}
\end{eqnarray}

\begin{eqnarray}
 W^{<}_{\alpha  k_\alpha \beta  n_\beta ,
       \gamma  k_\gamma \delta n_\delta}(\omega) &=&
\sum_{n_a , n_b} ~-~
 \frac{   K^{< ~ n_a  n_b}_{\alpha  k_\alpha \beta  n_\beta} \;
  \left(  K^{< ~ n_a  n_b}_{\gamma  k_\gamma \delta n_\delta} \right)^* }
     {\omega + \left( \varepsilon^{\pi}_{n_a}  +
             \varepsilon^{\pi}_{n_b} \right) - i \eta }
\nonumber \\
     &=&  {K^<}^\dag \frac{-1}{\omega -  E  - i \eta} {K^<}  \; ,
\label{eq:W<}
\end{eqnarray}
where $\varepsilon^{\pi}_{n_a}$ and $\varepsilon^{\pi}_{n_b}$ are the 
energies of the intermediate ph phonons.

\subsection{ERPA Matrix}
\label{app:A.ERPA.1}

The  eigenvalue equation for the ERPA is obtained in the usual way, by
substituting Eqs.~(\ref{eq:Pif>}) to~(\ref{eq:Pi<}) into
Eq.~(\ref{eq:2PiERPA})
and then extracting the residues of the ph poles $\varepsilon^{\pi}_{n}$.
The result is an eigenvalue equation  in terms of the vectors ${\cal Z}^{>}$
and~${\cal Z}^{<}$.

To linearize the problem, we introduce the following components
\begin{eqnarray}
X^{(1)}_{n_\alpha k_\beta} &\equiv&  
 \frac{1}{\omega - D + i \eta } ~ G ~
       \left\{ \left(V ~+~ W^>(\omega) \right)  \left({\cal Z}^>\right)^*
             ~-~ \left(V ~+~ H^{>,<}     \right)  \left({\cal Z}^<\right)^* \right\} \; ,
\nonumber \\
Y^{(1)}_{k_\alpha n_\beta} &\equiv&  
 \frac{1}{\omega + D - i \eta } ~ G^* 
       \left\{ \left(V ~+~ H^{<,>}     \right)  \left({\cal Z}^>\right)^*
             ~-~ \left(V ~+~ W^<(\omega) \right)  \left({\cal Z}^<\right)^* \right\} \; ,
\nonumber \\
X^{(2)}_{n_a n_b} &\equiv&  
 \frac{1}{\omega - E + i \eta } ~ 
    K^>   \left({\cal Z}^>\right)^*     \; ,
\label{eq:defXY12}
 \\
Y^{(2)}_{n_a n_b} &\equiv&  
 \frac{1}{\omega + E - i \eta } ~ 
        K^<   \left({\cal Z}^<\right)^*    \; ,
\nonumber
\end{eqnarray}
where $X^{(1)}$
and $Y^{(1)}$ represent the ph amplitudes  that  appear in the standard
(D)RPA equations~\cite{schuckbook,brand90} and $X^{(2)}$ and $Y^{(2)}$ are
the analogous two-phonon amplitudes introduced
by the ERPA approach.
The components $X^{(1)}$ and $Y^{(1)}$ are related to ${\cal Z}^{>}$
and~${\cal Z}^{<}$ respectively by
\begin{eqnarray}
\left({\cal Z}^>_{\alpha n_\alpha \beta k_\beta}\right)^* &=&  
  {\cal X}^{n_\alpha}_{\alpha}{\cal Y}^{k_\beta}_{\beta}
      ~ X^{(1)}_{n_\alpha k_\beta}
  ~=~ G^\dag ~ X^{(1)}
\nonumber \\
\left({\cal Z}^<_{\alpha k_\alpha \beta n_\beta}\right)^* &=&  
  {\cal Y}^{k_\alpha}_{\alpha}{\cal X}^{n_\beta}_{\beta}
      ~ Y^{(1)} _{k_\alpha n_\beta}
  ~=~ (G^*)^\dag ~ Y^{(1)}
\label{eq:Z_vs_XY}
\end{eqnarray}
Eqs.~(\ref{eq:2PiERPA}) can be put in the form of
a linear eigenvalue equation
\begin{equation}
  \omega ~ 
   \left( \begin{array}{c}
     X^{(1)} \\  X^{(2)} \\  Y^{(1)} \\  Y^{(2)}  \end{array} \right)
   ~=~ 
     {\bf M}
     ~  \left( \begin{array}{c}
     X^{(1)} \\  X^{(2)} \\  Y^{(1)} \\  Y^{(2)}  \end{array} \right)  \; ,
\label{eq:ERPAeig}
\end{equation}
where the matrix $\bf{M}$ is defined as
\begin{equation}
  {\bf M} ~=~ 
    \left[ \begin{array}{cccc}
    G~V~G^\dag ~+~ D    &  G {K^>}^\dag &  G [ V~+~H^{>,<}] (G^*)^\dag  & \\
     K^>  G^\dag       &       E       &                          &   \\ 
   -G^* [ V~+~H^{<,>}] G^\dag & & - G^* V (G^*)^\dag - D 
                                                       & G^*  {K^<}^\dag \\
      &   &   K^<  (G^*)^\dag    &   - E   
                           \end{array} \right]  \; .
\label{eq:ERPAmtx}
\end{equation}
The off-diagonal 2$\times$2 blocks in Eq.~(\ref{eq:ERPAmtx}) describe
diagrams in which the time direction of propagation is inverted. In the
present case the only non vanishing elements are the ones that involve the
inversion of a single ph state into a hp one, or vice versa.
 These correspond to the sum of
the first-order term $V$, which represents the kernel of the bare RPA, and
the more complex diagrams of Fig.~\ref{fig:ERPAflip} .
Blank spaces would in principle allow to include more complicated
contributions, that involve time inversion of two-phonon diagrams into
ph or hp configurations. These contributions are not expected to play a
relevant role for the present problem.
It must be noted that if the terms involving the matrix $G~{K}^\dag$ are
discarded in Eq.~(\ref{eq:ERPAmtx}), the components $X^{(1)}$ and $Y^{(1)}$
decouple from $X^{(2)}$ and $Y^{(2)}$.  In this case, Eqs.~(\ref{eq:2PiERPA})
reduce to the ph-DRPA one~(\ref{eq:DRPA}) and the matrix~(\ref{eq:ERPAmtx})
would take the form of the standard RPA matrix~\cite{schuckbook}.
The normalization condition is derived in the usual way, by extracting the 
contribution of order zero of the expansion around a given pole and by
employing the conjugate equation. One eventually obtains~\cite{brand90}
\begin{equation}
  \sum_{i=1,2}  \left( {X^{(i)}}^\dag \; X^{(i)} ~-~ {Y^{(i)}}^\dag \; Y^{(i)} 
  \right) = 1 \; ,
\label{eq:ERPAnrm}
\end{equation}
where the inner product of the vectors $X^{(i)}$ and $Y^{(i)}$ is implied.

\subsection{Matrix Elements for ph~ERPA}
\label{app:A.ERPA.2}

In the following we give the explicit  expression for the matrix
elements of Eq.~(\ref{eq:ERPAmtx}).
The contributions originating from the standard (D)RPA equation are~(here
and below, summations over repeated greek indices are understood)
\begin{eqnarray}
  \left( G \, V \, G^\dag \right)_{n_1 k_2 , n_3 k_4} ~&=&~ 
  {\cal X}^{n_1}_{\alpha}{\cal Y}^{k_2}_{\beta} 
   ~ V_{\alpha \nu , \beta \mu} ~ 
  \left(  {\cal X}^{n_3}_{\mu}{\cal Y}^{k_4}_{\nu} \right)^*  
    \; ,
 \nonumber \\
 \nonumber \\
   \left( G^* \, V \, (G^*)^\dag \right)_{k_1 n_2 , k_3 n_4} ~&=&~ 
  \left(  {\cal Y}^{k_1}_{\alpha}{\cal X}^{n_2}_{\beta} \right)^*
   ~ V_{\alpha \nu , \beta \mu} ~
 {\cal Y}^{k_3}_{\mu}{\cal X}^{n_4}_{\nu} 
    \; ,
 \nonumber \\
 \nonumber \\
  \left( G^* \, V \, G^\dag \right)_{k_1 n_2 , n_3 k_4} ~&=&~
  \left(  {\cal Y}^{k_1}_{\alpha}{\cal X}^{n_2}_{\beta} \right)^*
   ~ V_{\alpha \nu , \beta \mu} ~
  \left(  {\cal X}^{n_3}_{\mu}{\cal Y}^{k_4}_{\nu} \right)^*  \; ,
 \label{eq:DRPA-Bmtx} \\
 \nonumber \\
   \left( G \, V \, (G^*)^\dag \right)_{n_1 k_2 , k_3 n_4} ~&=&~
  {\cal X}^{n_1}_{\alpha}{\cal Y}^{k_2}_{\beta} 
   ~ V_{\alpha \nu , \beta \mu} ~ 
 {\cal Y}^{k_3}_{\mu}{\cal X}^{n_4}_{\nu}
   \; ,
 \nonumber
\end{eqnarray}
with the corresponding unperturbed ph energies
\begin{equation}
  D_{n_1 k_2} = diag \{  \varepsilon^+_{n_1}  -
                  \varepsilon^-_{k_2}  \}  \; .
\nonumber \\
\nonumber
\end{equation}

The interaction between two-phonon intermediate states and the ph ones is
given by
\begin{eqnarray}
 \left( K^> \, G^\dag \right)_{n^\pi_a n^\pi_b , n_1 k_2} &=&
   \frac{1}{2} ~
   \left\{
     \left[
       \left(  {X^{(1)}}^{~n^\pi_a}_{n_\mu k_\rho} ~
               {X^{(1)}}^{~n^\pi_b}_{n_1 k_\epsilon}  \right)^*  ~
          {\cal X}^{n_\mu}_{\mu}{\cal Y}^{k_\rho}_{\rho} 
          ~ V_{\mu \nu , \rho \epsilon} ~ 
         {\cal Y}^{k_\epsilon}_{\epsilon}
              \left(  {\cal Y}^{k_2}_{\nu} \right)^* 
         \right.  \right.
\nonumber \\
\nonumber \\
          & &- ~  \left.  \left.
         \left(  {X^{(1)}}^{~n^\pi_a}_{n_\mu k_\rho} ~
               {X^{(1)}}^{~n^\pi_b}_{n_\nu k_2}  \right)^*  ~
          {\cal X}^{n_\mu}_{\mu}{\cal Y}^{k_\rho}_{\rho} 
          ~ V_{\mu \nu , \rho \epsilon} ~ 
          \left( {\cal X}^{n_1}_{\epsilon} \right)^* 
               {\cal X}^{n_\nu}_{\nu} 
     \right]  \right.
\nonumber \\
\nonumber \\
          & & ~ ~ ~ ~  + ~  \left.
     \left[ n^\pi_a  \leftrightarrow n^\pi_b  \right]
   \right\}
\end{eqnarray}
where $n^\pi_a$ and $n^\pi_b$ are the quantum numbers of the two phonons that
form the intermediate state.  The quantities ${X^{(1)}}^{~n^\pi}_{n_1 k_2}$
are the forward-going amplitudes of the intermediate
phonons~(\ref{eq:defXY12}) or~(\ref{eq:Z_vs_XY}) and are obtained
from the previous solution  of the polarization propagator.
 Note that $K^>G^\dag$ is symmetric under the exchange of the two indices
$n^\pi_a$ and $n^\pi_b$, as required by the boson character of the
ph phonons.
 The factor $\frac{1}{2}$ assures that no double counting happens 
when only two free propagators~$\Pi^f(\omega)$ are coupled.
 Due to the time-inversion properties of npnh states, the corresponding
contribution for backward-going propagation is simply related to above one:
\begin{equation}
 \left( K^< \, (G^*)^\dag \right)_{n^\pi_a n^\pi_b , k_1 n_2} 
 ~=~  \left\{
 \left( K^> \, G^\dag \right)_{n^\pi_a n^\pi_b , n_2 k_1}  \right\}^* \; .
 \nonumber
\end{equation}

The analytical expression for the contribution of the time-inversion diagrams
is a little more complicated.
 First, we introduce the following quantity that corresponds to the 
generation of a two-phonon state by a fluctuation of the nuclear mean
field
\begin{equation}
  U_{n^\pi_a n^\pi_b} =
   \frac{1}{- \left( \varepsilon^\pi_{n_a}  +
                  \varepsilon^\pi_{n_b} \right)} ~
     \left[
         \left( {\cal Y}^{k_\sigma}_{\sigma}
                {\cal Y}^{k_\lambda}_{\lambda} \right)^* 
           V_{\sigma  \lambda , \mu \nu}  
         \left( {\cal X}^{n_\mu}_{\mu} 
                {\cal X}^{n_\nu}_{\nu} \right)^*  ~
        {X^{(1)}}^{~n^\pi_a}_{n_\mu k_\sigma}
        {X^{(1)}}^{~n^\pi_b}_{n_\nu k_\lambda}  \; . 
      \right]
\label{ERPA-Bmtx1}
\end{equation}
This is also symmetric under the exchange of the indices $n^\pi_a$ and~$n^\pi_b$.
Then, the matrix elements containing the time-inversion diagrams,
$H^{>,<}$ and~$H^{<,>}$, can be written as follows (a summation over $n^\pi_a$
and~$n^\pi_b$ is also implied)
\begin{eqnarray}
 \left( G \, H^{>,<} \, (G^*)^\dag \right)_{n_1 k_2 , k_3 n_4} ~&=&~
    \left\{ 
    \left( (G^*) \, H^{<,>} \, G^\dag \right)_{k_3 n_4 , n_1 k_2} 
  \right\}^* ~=~
\nonumber \\
   &=&~ \frac{1}{2} 
    \left\{
     \left[
          {\cal X}^{n_1}_{\alpha} {\cal Y}^{k_2}_{\beta} 
          ~ V_{\alpha \epsilon , \beta \rho} ~ 
         \left(  {\cal X}^{n_\rho}_{\rho} 
             {\cal Y}^{k_\epsilon}_{\epsilon}  \right)^* 
               {X^{(1)}}^{~n^\pi_a}_{n_\rho k_3} ~
               {X^{(1)}}^{~n^\pi_b}_{n_4 k_\epsilon}    ~
         \right]  \right.
\nonumber \\
\nonumber \\
          & & ~ \left. 
    ~ ~ ~ ~ ~ + ~ \left[ - \left( n_1  \leftrightarrow n_4 \right)
             ~ - \left( k_2  \leftrightarrow k_3 \right)
             ~ + \left(n_1  \leftrightarrow n_4  , k_2  \leftrightarrow k_3 
             \right)  \right]
    \right.
\nonumber \\
\nonumber \\
          & & ~ \left. 
       + ~ \left[
          {\cal X}^{n_1}_{\alpha} {\cal Y}^{k_3}_{\gamma} 
          ~ V_{\alpha \epsilon , \gamma \rho} ~ 
         \left(  {\cal X}^{n_\rho}_{\rho} 
             {\cal Y}^{k_\epsilon}_{\epsilon}  \right)^* 
               {X^{(1)}}^{~n^\pi_a}_{n_\rho k_\epsilon} ~
               {X^{(1)}}^{~n^\pi_b}_{n_4 k_2}    ~
         \right]  \right.
\nonumber \\
\nonumber \\
          & & ~ \left. 
        ~ ~ ~ ~ ~ ~ ~ ~ ~ ~ ~ ~ ~ + ~ 
        \left[ n_1  \leftrightarrow n_4  , k_2  \leftrightarrow k_3 
              \right]
    \right.
\nonumber \\
\nonumber \\
          & & ~ \left. 
       + ~  {\cal X}^{n_1}_{\alpha} {\cal X}^{n_4}_{\delta}
          ~ V_{\alpha \delta , \rho \lambda} ~ 
         \left(  {\cal X}^{n_\rho}_{\rho} 
             {\cal X}^{n_\lambda}_{\lambda}  \right)^* 
               {X^{(1)}}^{~n^\pi_a}_{n_\rho k_2} ~
               {X^{(1)}}^{~n^\pi_b}_{n_\lambda k_3}   ~
       \right.
\nonumber \\
\nonumber \\
          & & ~ \left.
        + ~    {\cal Y}^{k_\rho}_{\rho}  {\cal Y}^{k_\lambda}_{\lambda} 
          ~ V_{\rho \lambda , \beta \gamma} ~ 
         \left(  {\cal Y}^{k_2}_{\beta} 
             {\cal Y}^{k_3}_{\gamma}  \right)^* 
               {X^{(1)}}^{~n^\pi_a}_{n_1 k_\rho} ~
               {X^{(1)}}^{~n^\pi_b}_{n_4 k_\lambda}   ~
   \right\} \left( U_{n^\pi_a n^\pi_b} \right)^*   \; .
\label{ERPA-Bmtx2}
\end{eqnarray}

Finally, the two-phonon unperturbed energies are given by
\begin{equation}
  E_{n^\pi_a n^\pi_b} = diag \{  \varepsilon^\pi_{n_a}  +
                  \varepsilon^\pi_{n_b}  \}  \; .
\end{equation}




\begin{table}
 \begin{center}
 \begin{tabular}{ccccccccccc}
  \hline 
  \hline 
 $T=0$
  & \hspace{.1truein}   & \multicolumn{2}{c}{h.o./RPA} 
  & \hspace{.1truein}   & \multicolumn{5}{c}{h.o./ERPA} 
  &  \\
 $J^\pi$  &  & $\varepsilon_{n_\pi}$ & $Z_{n_\pi}$ 
          &  & $\varepsilon_{n_\pi}$ & $Z_{n_\pi}$
          & \hspace{.05truein} & $ph$(\%) & $2\Pi$(\%) 
          & \\
  \hline 
  \hline 
 $2^+$ &  &  22.86 &   1.039 &    &   23.13 &    0.823  & & 79    & 21    &   \\
 $2^+$ &  &        &         &    &   21.30 &    0.133  & & 13    & 87    &   \\
 $2^+$ &  &        &         &    &   19.11 &    0.118  & & 11    & 89    &   \\
   \\
 $0^+$ &  &        &         &    &   14.28 &    0.010  & &  1    & 99    &   \\
 $2^+$ &  &        &         &    &   13.91 &    0.041  & &  4    & 96    &   \\
 $4^+$ &  &        &         &    &   13.87 &    0.028  & &  3    & 97    &   \\
   \\
 $1^-$ &  &   9.13 &   1.027 &    &    9.20 &    1.075  & & 99.7  &  0.3  &   \\
 $3^-$ &  &   7.14 &   1.258 &    &    6.55 &    1.269  & & 97    &  3    &   \\
 $0^+$ &  &   9.46 &   1.582 &    &    6.52 &    1.820  & & 80    & 20    &   \\
  \hline 
  \hline 
 \end{tabular}
  \end{center}
    \parbox[t]{1.\linewidth}{
      \caption[Two-phonon ERPA spectrum and strengths; h.o. and BHF inputs]
  {  \small 
    Excitation energy and total spectral strengths obtained for the 
   principal solutions of the RPA and ERPA equations.
    For the ERPA case the total fraction of ph and two-phonon contributions
   are also shown.
    An IPM input sp propagator was used to generate these results.
    \label{tab:2PiERPA_ho}  }
    }
\end{table}

\begin{table}
 \begin{center}
 \begin{tabular}{ccccccccccccccc}
  \hline 
  \hline 
 $T=0$
   & \hspace{.1truein}   & \multicolumn{2}{c}{dressed/DRPA} 
   & \hspace{.1truein}   & \multicolumn{5}{c}{dressed/ERPA}
          & $(0^+)^2$   & $(3^-)^2$
          & $(0^+,3^-)$ & $(0^+,1^-)$
          & \\
 $J^\pi$  &  & $\varepsilon^\pi_n$ & $Z_{n_\pi}$ 
          &  & $\varepsilon^\pi_n$ & $Z_{n_\pi}$
          & \hspace{.05truein} & $ph$(\%) & $2\Pi$(\%) 
          & (\%)   & (\%)  & (\%)   & (\%)
          & \\
  \hline 
  \hline 
 $2^+$ &  &  23.77 &   0.468 &   &  23.52 &  0.123 &   & 26   & 74   & \\
 $2^+$ &  &        &         &   &  22.96 &  0.341 &   & 78   & 22   & \\
 $2^+$ &  &  20.59 &   0.269 &   &  20.42 &  0.255 &   & 98   &  2   & \\
       \\
 $1^-$ &  &        &         &   &  13.37 &  0.148 &   & 21   & 79   &      &      &      &  79   & \\
 $3^-$ &  &        &         &   &  12.35 &  0.113 &   & 16   & 84   &      &      & 84    \\
       \\
 $0^+$ &  &        &         &   &  12.15 &  0.001 &   &  1   & 99   &  3   & 96   \\
 $4^+$ &  &        &         &   &  12.14 &  0.007 &   &  1   & 99   &      & 99   \\
 $2^+$ &  &        &         &   &  12.12 &  0.008 &   &  1   & 99   &      & 98   \\
       \\
 $0^+$ &  &  16.62 &   0.717 &   &  17.21 &  0.633 &   & 88   & 12   &  10  &  0.5 \\
       \\
 $0^+$ &  &        &         &   &  11.28 &  0.092 &   & 12   & 88   &  85  &  2   \\
       \\
 $1^-$ &  &  11.19 &   0.720 &   &  10.90 &  0.680 &   & 94.1 &  5.9 &      &      &      &  5.8  & \\
 $3^-$ &  &   9.50 &   0.762 &   &   9.23 &  0.735 &   & 95.9 &  4.1 &      &      &  4.0  \\
  \hline 
  \hline 
 \end{tabular}
  \end{center}
    \parbox[t]{1.\linewidth}{
      \caption[Two-phonon ERPA spectrum and strengths; dressed input]
  {  \small 
    Excitation energy and total spectral strengths obtained for the 
   principal solutions of DRPA and two-phonon ERPA equations. A
   dressed sp propagator was employed.
    The total contribution of ph and two-phonon states
   of the ERPA solutions are shown.
    For states below 15~MeV, the columns on the right side give the 
individual contributions of
all the relevant two-phonon contributions. The sum
of these terms for the states that are listed does not
   exceed 1\%.
    \label{tab:2PiERPA_3g}  }
    }
\end{table}

\begin{table}
 \begin{center}
 \begin{tabular}{cccccccccccccccccccc}
  \hline 
  \hline 
 $T=0$
  & \hspace{.2truein}    & \multicolumn{2}{c}{h.o./ERPA} 
  & \hspace{.1truein}    & \multicolumn{2}{c}{h.o./ERPA} 
  & \hspace{.1truein}    & \multicolumn{2}{c}{h.o./ERPA} 
  & \hspace{.3truein}    & \multicolumn{2}{c}{dressed/ERPA} 
  & \hspace{.05truein}   & \multicolumn{2}{c}{dressed/ERPA}
  & \hspace{.05truein}   & \multicolumn{2}{c}{dressed/ERPA}
  &  \\
  &  & & &  &\multicolumn{2}{c}{no time-inv.} & & \multicolumn{2}{c}{$\rho$=5 mod. sp.}
  &  & & &  &\multicolumn{2}{c}{no time-inv.} & & \multicolumn{2}{c}{$\rho$=5 mod. sp.} & \\
 $J^\pi$  &  & $\varepsilon_{n_\pi}$ & $Z_{n_\pi}$ 
          &  & $\varepsilon_{n_\pi}$ & $Z_{n_\pi}$
          &  & $\varepsilon_{n_\pi}$ & $Z_{n_\pi}$
          &  & $\varepsilon_{n_\pi}$ & $Z_{n_\pi}$
          &  & $\varepsilon_{n_\pi}$ & $Z_{n_\pi}$ 
          &  & $\varepsilon_{n_\pi}$ & $Z_{n_\pi}$
          &   \\
  \hline 
  \hline 
 $2^+$ &  &   23.13 &    0.823  & &  23.09 &   0.824 & &        &         &     &   22.96 &    0.341 &     &  22.96 &   0.342 &     &   22.39 &    0.210 &  \\
 $2^+$ &  &         &           & &        &         & &  19.25 &   0.441 &     &   21.44 &    0.163 &     &  21.43 &   0.161 &     &   21.35 &    0.122 &  \\
 $2^+$ &  &         &           & &        &         & &  17.13 &   0.471 &     &   20.42 &    0.255 &     &  20.41 &   0.258 &     &   20.30 &    0.281 &  \\
   \\
 $0^+$ &  &   14.28 &    0.010  & &  14.28 &   0.010 & &  12.85 &   0.232 &     &   12.15 &    0.001 &     &  12.15 &   0.001 &     &   12.06 &    0.005 &  \\
 $2^+$ &  &   13.91 &    0.041  & &  13.91 &   0.041 & &  13.88 &   0.101 &     &   12.12 &    0.008 &     &  12.12 &   0.008 &     &   12.05 &    0.013 &  \\
 $4^+$ &  &   13.87 &    0.028  & &  13.87 &   0.028 & &  11.81 &   0.036 &     &   12.14 &    0.007 &     &  12.14 &   0.007 &     &   12.09 &    0.009 &  \\
   \\
 $0^+$ &  &         &           & &        &         & &        &         &     &   11.28 &    0.092 &     &  11.28 &   0.092 &     &   10.97 &    0.184 &   \\
   \\
 $1^-$ &  &    9.20 &    1.075  & &   9.03 &   1.024 & &   9.21 &   1.260 &     &   10.90 &    0.680 &     &  10.89 &   0.680 &     &   10.64 &    0.723 &  \\
 $3^-$ &  &    6.55 &    1.269  & &   6.54 &   1.266 & &   4.68 &   1.581 &     &    9.23 &    0.735 &     &   9.22 &   0.735 &     &    8.91 &    0.762 &  \\
 $0^+$ &  &    6.52 &    1.820  & &   6.36 &   1.596 & &   6.74 &   3.055 &     &         &          &     &        &         &     &         &          &  \\
  \hline 
  \hline 
 \end{tabular}
  \end{center}
    \parbox[t]{1.\linewidth}{
      \caption[no time inversions and larger model space; h.o. and BHF inputs]
  {  \small 
    The excitation energy and total spectral strengths discussed in
Sec.~\ref{sec:results} are compared to the solutions of the two-phonon
    ERPA equations by neglecting the time-inversion diagrams
    of Fig.~\ref{fig:ERPAflip}.
     Both IPM and dressed input propagators cases are displayed.
     Results obtained within a larger model space ($\rho$=$2n+l$=5)
    are also shown.
    \label{tab:B_mtx}  }
    }
\end{table}

\end{document}